\documentclass[aps,pre,twocolumn,groupedaddress]{revtex4-1}

\usepackage{graphicx}

\begin{document}



\title{Light enhancement on thin and ultra-thin high-index dielectric slabs with  rectangular nano-pits}


\author{J\'er\^ome Le Perchec}
\email[]{jerome.le-perchec@cea.fr}

\affiliation{CEA, LETI, Minatec Campus, Optics and Photonics Department, 17 avenue des Martyrs, 38054 Grenoble, France}

\begin{abstract}
We closely study the local amplifications of visible light on a thin dielectric slab presenting a sub-wavelength array of small, rectangular, bottom-closed holes. The high-quality Fabry-Perot resonances of eigen modes which vertically oscillate, and their corresponding near-field maps, especially inside the voids, are numerically quantified with RCWA and analytically interpreted through a quasi-exact modal expansion. This last method gives explicit opto-geometrical rules allowing to finely understand the general trends in 1D and 2D. In more advanced examples, we show that multi-cavity and/or slightly thicker two-dimensional gratings may generate anomalously frequency-susceptible surfaces over a broad spectral range. Also, dielectric membranes a few nanometers thick only, can catch light, with tremendous enhancements of the electric field intensity ($>10^6$) that largely extends in the surrounding space. 
\end{abstract}


\maketitle

\section{Introduction}

Sub-wavelength gratings offer a great richness of optical properties for filtering, absorption and local amplification of light coupled in free-space, and, depending on whether the material is dielectric or metallic, allow getting different behaviors to address wanted specifications \cite{Collin}. 
All-dielectric metamaterials and metasurfaces have especially met a renewed and increasing attention from the photonics community, in the current decade, because of their lossless optical properties and a greater plurality of electrical and magnetic modes \cite{Kuznetsov}, by comparison to surface-plasmon-based structures, especially regarding efficient Mie or Fano resonances \cite{Staude} and the so-called bound states in the continuum \cite{Hsu2013}. 

In the case of planar-type structures, the existence of abrupt spectral resonances in dual dielectric layers or high-optical-index gratings is known since a long time \cite{Botten1981,Bertoni}. They have been continuously studied on two-dimensional (2D) photonic crystal slabs \cite{Peng,Fan2002} and, more recently, on strong index-contrast gratings \cite{ChangHasnain} which indeed present relevant applications in opto-electronics, because of the high quality(Q)-factors they allow. The resonant spectral properties of dielectric grating-waveguides have been extensively used \cite{Rosenbatt,Tibuleac2000,Coves2004,Fehrembach2007} to make selective reflection or transmission band-pass filters, ranging from microwave to optical frequencies. 
The near-field resonances were studied to design, for instance, photo-active substrates and bio-sensors \cite{Ganesh,Chaudhery,Beheiry,Nicolaou}, or to efficiently generate second or third harmonic waves \cite{Ban2019}. 

\begin{figure}
\centering
\includegraphics[angle=0,scale=.37]{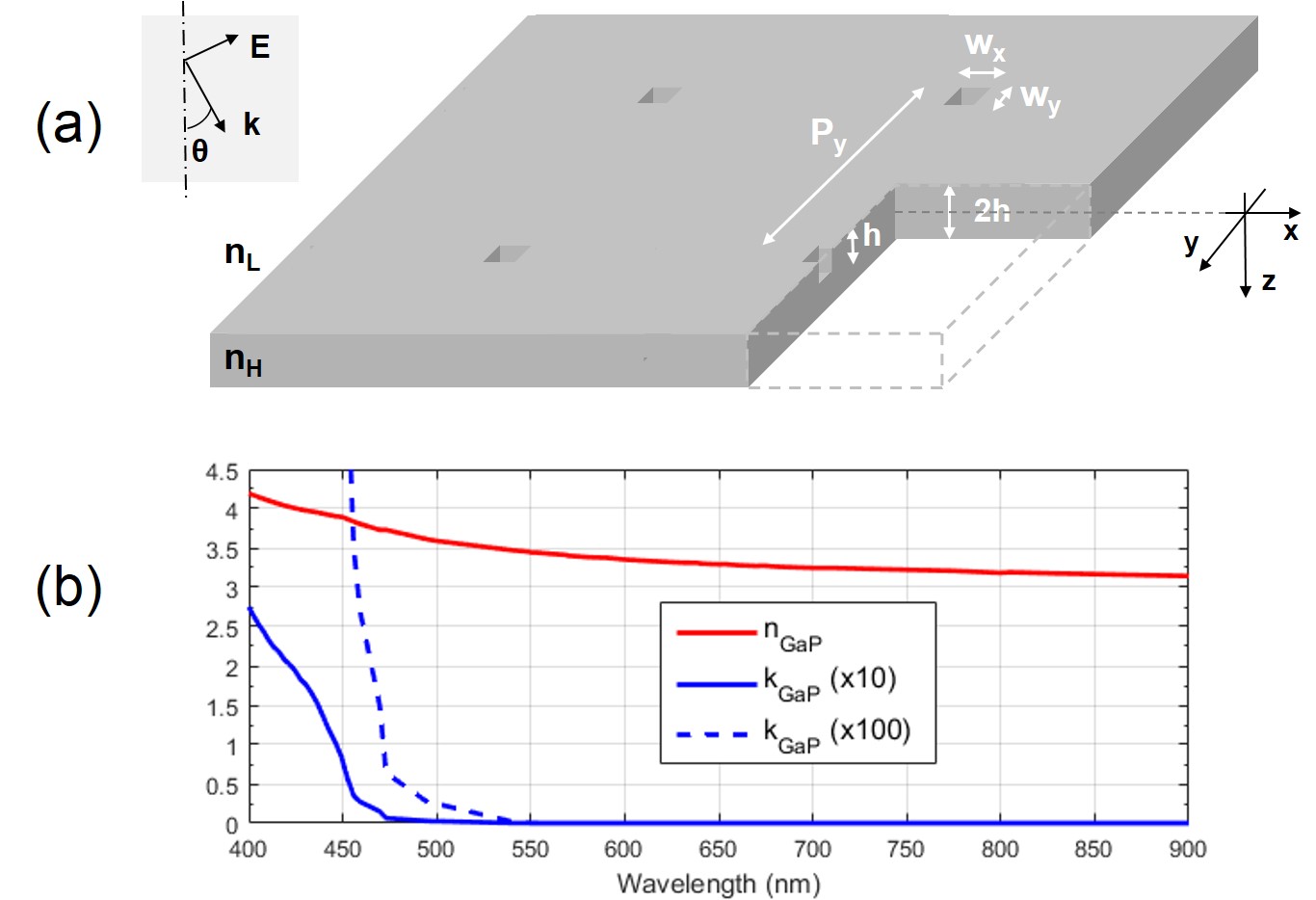}
\caption{(a) 2D grating with rectangular bottom-closed holes made a high-index dielectric plate. The cavity depth ($h$) is only half that of the slab. The external low index medium is air ($n_L=1$). Light comes from above. The $(x,z)$-plane is the incidence plane, containing the (\textbf{k,E}) vectors ($k=2 \pi/\lambda$) in TM-polarization, and being orthogonal to the incident magnetic field \textbf{H}. Angle $\theta$ is defined between \textbf{k} and $z$-axis. (b) Optical index and extinction coefficient $(n,k)$ of undoped Gallium Phosphide over the visible range (values taken from \cite{Palik}).}\label{figure1}
\end{figure}

Regarding light localization in small accessible volumes (of subwavelength scale), plasmonic structures had attracted a preponderant interest by the past, but lossless all-dielectric platforms, like the ones based on Fano resonances (possibly tunable \cite{Zhou}), have come back to the front-stage \cite{Chang2018,Staude} and are complementary to non-plasmonic nano-antennas \cite{Caldarola}. Sharp Fabry-Perot resonances occurring on thin high-index slabs, containing nano-trenches, may lead to local electric-field amplifications of several orders of magnitude, as high as that obtainable with surface plasmons in metallic grooves \cite{LePerchec19}. Also, a continuous or all-dielectric waveguide layer is not mandatory to get good resonances: metallic walls can be inserted inside the unit cell provided the dielectric regions have a minimum \textit{lateral} size to support a vertically propagating mode. Thus, managing the light confinement and amplification in resonant holes or slots made in dielectric slabs (especially in the visible spectrum where high-index materials are scarce) remains a subject of prime interest, for which many further works are still expected.


 In this paper, following the study recently reported \cite{LePerchec19}, we present a generalized analysis of the more complex 2D nano-pit grating, which consists in a thin, high dielectric-index slab with bottom-closed voids on one side, and here illuminated by visible light. Contrary to photonic crystals like honey-comb structures \cite{Notomi} 
 with classical circular holes, we consider rectangular cavities, like the ones sketched in Fig.\ref{figure1}(a), occupying a very small fraction of the slab volume. Their miniaturization is \textit{essential} to get strong fields inside. Based on a reference example with Gallium Phosphide (wide band-gap semiconductor), we observe resonances with Q-factors much higher than in 1D, and over-intensities of the electric field $\sim 10^5-10^6$ within the voids, which are not reachable with metallic structures of same geometry (sections \ref{section:refexample} and \ref{section:maps}). Transparent substrates containing small pits may constitute a solution useful for Raman spectroscopy or fluorescence enhancement, and, depending on polarization states, spatial maps clearly indicate how the electromagnetic field localizes. A very special effort was done, in the intermediate section \ref{section:modaldev}, to give explicitly the key parameters allowing to describe qualitatively and quantitatively the general trends of the photonic responses both in 1D and 2D, thanks to a \textit{true} modal approach. 
 We notably point out (sections \ref{section:freqsusceptibility} and \ref{section:ultrathin}) the possibility of getting resonant surfaces with more irregular patterns and showing an anomalous frequency-susceptibility over a broad visible range, i.e. sometimes a great spectral density of strong intensity peaks (up to one resonance per nm wavelength). Also, ultra-thin membranes $(\sim \lambda/100)$ can support giant resonances with maximum spreading of the enhanced field in the empty space surrounding the slab (section \ref{section:ultrathin}).

\section{Simulations of a reference example}
\label{section:refexample}
All numerical simulations were carried out through Rigorous Coupled Wave Analysis (RCWA) adapted for bi-dimensional gratings \cite{Hazart}. As we investigate high Q-factor resonances, spectra were calculated with a wavelength resolution of at least $0.01$nm, and a bi-perodic modal basis of more than $30\times 30$ Fourier orders is used for near-field mapping, which is quite time-consuming for computation. 

By default, we consider a TM excitation wave whose \textbf{H} field is orthogonal to the incidence plane, as defined in Fig.\ref{figure1}(a). In such a configuration, it is expected to excite pure TM eigen modes, and quasi-TE grating modes (in fact, hybrid TM/TE modes) since the resonators virtually see one of both linear polarizations along each orthogonal periodic direction. Special attention will be paid regarding the electric field enhancements.

\begin{figure}
\centering
\includegraphics[scale=.41]{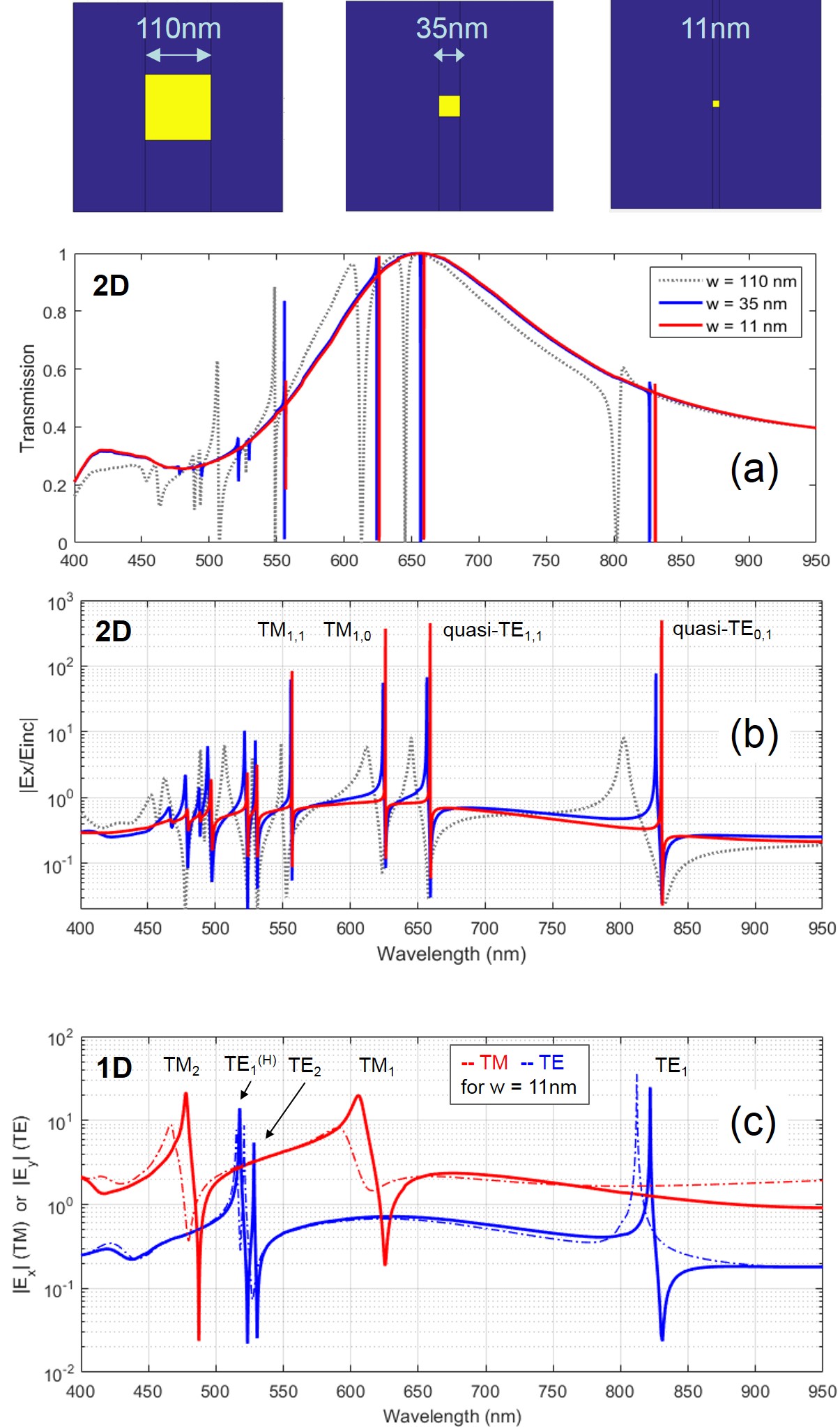} 
\caption{(a) Transmission spectra of the grating described in Fig.~\ref{figure1}(a), under TM-excitation, with $n_H=n_{GaP}$, $P_{x,y}=P=350$nm, taking pits of different sizes $w_{x,y}=w=110$, $35$ and $11$nm, at normal incidence. The area ratio $R=w^2/P^2$ varies from 10 to 1012. The structure thickness is $2h=100$nm (void depth $h=50$nm). (b) Spectra of the normalized $E_x$ component in logarithmic scale, calculated at \textit{mid-height} of the 2D cavities at $(x,y,z)=(P_x/2,P_y/2,-h/2)$. Each FP resonance peak is associated to a dominating eigen mode according to the notations described in Section III. (c) Comparison to a 1D grating ($w_y=P_y=\infty$) for each linear polarization, with $w=11$nm. Solid lines: groove grating (bottom-closed cavities); dash-dotted lines: open slit grating (field still calculated at $z=-h/2$). Upper-script (H) means harmonic resonance.}\label{figure2}
\end{figure}
We exemplify this work in the visible range with a Gallium Phosphide slab whose high refractive index ($n_H>3$) and extinction coefficient are recalled in Fig.\ref{figure1}(b). GaP becomes optically absorbing when $\lambda<560nm$ (indirect band-gap semiconductor). This material shows interesting properties for optical systems working in the visible or infrared ranges \cite{Vaclavik}. Let us consider a square grating $100$nm thick, with a sub-wavelength period $P_{x,y}=P=350nm$, and three hole sizes $w_{x,y}=w=110$, $35$ and $11$nm, each one characterized by the respective ratios $R=(P/w)^2\approx 10$, $10^2$ and $10^3$.

Figures \ref{figure2}(a) and (b) display the spectra of the far-field transmission and the electric field amplification ($E_x$ component here) inside the nano-pits. We immediately see that the far-field shows a broad transmission peak typical of a homogeneous dielectric slab, marked by several sharp dips leading to zero transmission (and total reflection, due to energy conservation). This structure could play the role of an optical band-pass filters, with selectively rejected wavelengths (the latter being finely adjustable with the incidence angle). The transmission dips are correlated to very strong intensity enhancements $|E_x/E_{inc}|^2$ inside the voids (from $10^2$ to $10^5$ typically, calculated at mid-height of the cavities). As $R$ decreases, the resonance spectral widths clearly reduces whereas the field enhancement increases. We get the most sub-wavelength case at $\lambda (nm)=830.76=5.24n_Hh$ when $w=11nm$ (i.e. $\lambda/w=75.5$ and $\lambda/h=16.6$). This corresponds to $|E_x/E_{inc}|^2 \approx 2.5.10^5$ at $z=-h/2$, and a quality factor $Q > 10^5$. However, compared to the ratio $R=10$, resonances with $R=10^3$ are more damped, or even killed, when GaP becomes absorbing at shorter wavelengths. The quality factor may be estimated as $Q=\lambda_{res}/\Delta \lambda$ and also means:
\begin{equation}
Q=(2\pi)\frac{\text{Stocked\ power}}{\text{Dissipated\ (radiated\ or\ absorbed)\ power}}.
\end{equation}

In absence of any absorption, simulations show that it is essentially governed by the hole surface (at fixed period) i.e. by the radiation leaks escaping from the voids through the zero-order diffraction beam, which is confirmed by analytic development in the next section. With voids a few nanometers wide only, the 2D grating tends towards an almost perfect resonator stocking maximum energy. We also remark in Fig.\ref{figure2}(b) that, around each resonance, the field amplitude always meets a zero and a pole, which come closer when $w$ decreases. For a given aperture size, taking the square case $w_x=w_y=35$nm or the case $w_x=50$ and $w_y=20$nm for instance, does not modify substantially the quantitative results (the rectangular case may lead to a slightly higher enhancement).

By comparison, simulations of an equivalent 1D grating (taking $w_y=P_y=\infty$) is shown in Fig.\ref{figure2}(c). We get much lower intensities $|E_x|^2\sim 20^2$ (at mid-height of cavity) even with the narrower trench case $w_x=11$nm, and whatever the incident polarization, because of a greater aperture in the $y$-direction. We also note that open slits are clearly less favorable to local enhancements, especially at TM-resonances (we will explain it later). The normalized field modulus rapidly falls under the threshold of 10 with $w=35$nm (not shown).

These phenomena are all vertical FP-type resonances of different bi-periodic eigen modes inside the high optical index regions, which yield over-intensities inside narrow empty cavities. The dominating mode at each enhancement peak has been noted $TM_{m,p}$ or quasi-$TE_{m,p}$ in Fig.\ref{figure2}(b), in accordance with field maps displayed in Figures \ref{figure4} and \ref{figure5}. These maps will be described later in order to explain, first, the notations used through a very instructive modal analysis. This is the object of the following important section.

\section{True modal expansion and key parameters of the sharp resonances}
\label{section:modaldev}

A dielectric grating may support a range of  propagating or evanescent modes which can all store some energy from the incident excitation wave. It could seem difficult to extract simple analytic expressions allowing to describe quantitatively the resonance phenomena that are numerically obtained and to identify which modes or which key parameters are involved. It is however possible by resorting to the exact modal method \cite{Botten1981} which uses the true eigen modes, and not Fourier modal expansions (as is the case in RCWA). The method is very talking on the physical level and is well established in 1D, but not in 2D. Also, it needs a careful mathematical treatment to solve the problem in case of absorbing materials. With dielectrics, the method is simpler, and the case of small cavities (compared to the period) is much adapted to find explicit formulas. 
We will first explain the modal behaviour of the 1D sub-$\lambda$ lamellar grating, especially at normal incidence. By this, way, the bi-periodic structure containing small rectangular holes can be tackled in a more easy way, without giving exhaustive demonstrations.

\subsection{1D grating (TM or TE case)}
\begin{figure}
\begin{center}
\includegraphics[angle=0,scale=.49]{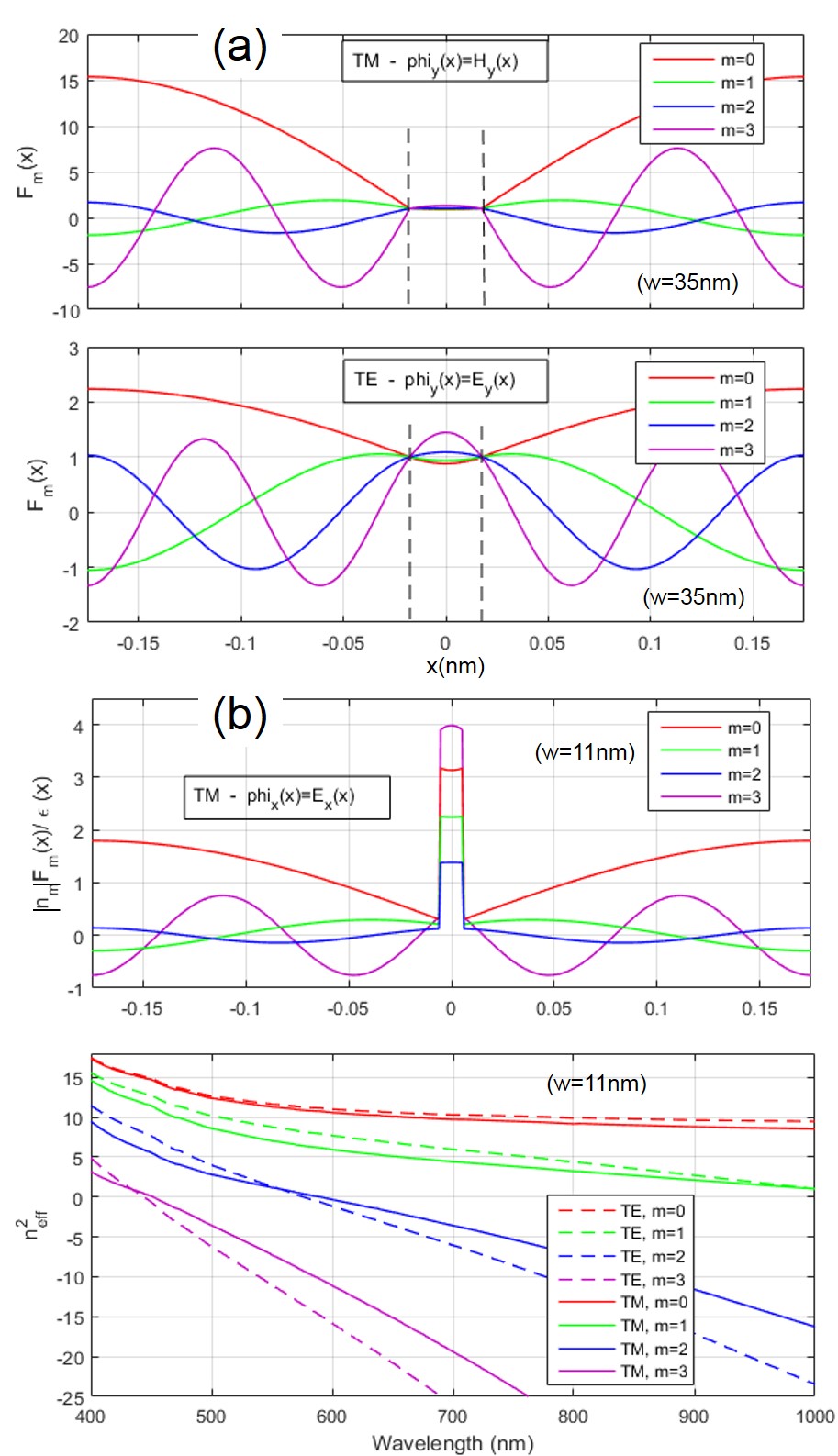} 
\end{center}
\caption{1D case. (a) Exact eigen functions $F_m(x)$ of the even modes which can be excited at normal incidence and at $\lambda=650$nm, for a GaP lamellar grating characterized by $P=350$nm and $w=35$nm, both for TM and TE polarizations (see Eq.(\ref{phiy})). The origin of abscissa is here taken at the center of the cavity and all $F_m$ are normalized to 1 at the vertical walls. (b) Example with $w=11$nm, in TM-polarization: normalized eigen mode $E_x^m(x)$ (see Eq.(\ref{phix})) at $\lambda=650$nm, and corresponding effective indices (square values) over the visible range (exact solutions of Eq.(\ref{eqtranscendentalTMl})). If $n^2_{eff,m}<0$, the mode is said evanescent in the $z$-direction (no possible FP resonance) and describes a rapidly oscillating surface wave at the horizontal interfaces. When GaP becomes absorbing at short wavelengths, $n^2_{eff}$ develops an imaginary part we have omitted here. In TM, $E_x$ always show an intrinsic amplitude jump within the narrow cavity (whether the mode resonates or not). In TE, fields are continuous at the vertical interfaces.}
\label{figure3}
\end{figure}

We consider the simple bar grating ($w_y$ and $P_y \rightarrow \infty$) whose total thickness is $2h$, surrounded by air. One can show that the field inside the grating reads: 
\begin{eqnarray} \label{phiy}
\phi_y= \sum_m F_m(x) [A_m\cos(kn_{m}z)+B_m\sin(kn_{m}z)], \quad \\
\phi_x= \sum_m \frac{i n_m}{\omega(x)}F_m(x)[-A_m\sin(kn_{m}z)+B_m\cos(kn_{m}z)]  \label{phix},\quad
\end{eqnarray}
where $\phi_y$ is the magnetic (respectively electric) field in TM-polarization (respectively TE-case) and $\phi_x=E_x$ (respectively $H_x$) in TM (respectively TE) polarization. $A_m$ and $B_m$ are complex excitation coefficients.
$F_m$ is the normalized eigen function of the $m^{th}$ grating mode which describes the field behaviour $\phi_y$ in function of x (see example in Fig.\ref{figure3}(a)). These functions constitutes a rigorously orthogonal basis, and are classically of the form $a\cos(k_x x)+b\sin(k_x x)$ in each dielectric region of the structure. The effective indices $n_{eff,m}=n_m$, are solutions of a well-known transcendental equation  \cite{Botten1981} we do not recall here. The weight function $\omega(x)$, in (\ref{phix}), is:
\begin{equation}
 \omega(x)=\varepsilon(x) \text{ in TM, }\omega(x)=1 \text{ in TE.} 
\end{equation}
It naturally derives from Maxwell equations and has importance in the coupling terms (see hereafter).
Outside the grating, the transmitted field is diffracted in Rayleigh orders which forms the P-periodic basis $(e_q)_q=(e^{i k \gamma_q x})_q$, with $\gamma_q=\sin\theta +q\lambda/P$:
\begin{equation} 
\phi_y(x,z>h)= \sum_{q=-\infty}^{+\infty} T_q e^{i k( \gamma_q x+\beta_q (z-h))},
\end{equation} 
where $\beta_q=\sqrt{1-\gamma_q^2}$, by taking $n_L=1$ to simplify.
 
Given appropriate continuity conditions and orthogonality rules of the eigen bases, we judiciously apply the method of moments \cite{Botten1981,Suratteau} and get a matrix solution giving unambiguously the coefficients $A_m$ and $B_m$. In this work, we considerably simplify the matrix solving by resorting to a \textit{weak  indirect-coupling} condition between the eigen modes (see explanation in Appendix A, section \ref{appendixA}). By this way, it is possible to extract $A_m$ and $B_m$, for each mode independently:
\begin{eqnarray}
\label{Am}
A_m= \frac{<F_m|e_0>}{\cos(kh n_{m})W_m -  i n_{m}\sin(kh n_{m}) \sum_q \frac{ |<e_q|F_m>|^2}{\beta_q} },\quad  \\
\label{Bm}
B_m = \frac{<F_m|e_0>}{\sin(kh n_{m})W_m + i n_{m}\cos(kh n_{m}) \sum_q \frac{ |<e_q|F_m>|^2}{\beta_q} }.\quad 
\end{eqnarray}
This is a quasi-exact general expression for modes with expected high Q-factors in the all-dielectric grating, where the coupling terms (overlap integrals) are:
\begin{equation} \label{coupling}
<f|g>=\frac{1}{P} \int_{0}^{P} f^* (x)\frac{1}{\omega(x)} g(x) dx.
\end{equation} 

$W_m=<F_m|F_m>$ can be viewed as a norm of the eigen functions. Generally, each $F_m$ does not change of profile and amplitude throughout the spectrum of study while the permittivity remains nearly constant, i.e $W_m$ weakly varies with frequency.

At normal incidence, only \textit{even} eigen-functions $F_m$ can be excited and are almost unitary in a narrow cavity (see Fig.\ref{figure3}(a)) whereas $F_m(x)\sim \cos(k_x x)$ in the high-index regions, except for the fundamental mode $m=0$ which does not present field node. At view of these oscillating behaviours, $W_0$ is a few unities, whereas $W_m\sim 0.5 $ for the first higher-order modes (this is not necessarily true for all modes of the eigen basis). 
The effective indices of even modes are solution of the dispersion relation:
\begin{equation}
\label{eqtranscendentalTMl}
 \tan \left( \frac{k_x^L w_x}{2} \right) + \frac{1}{Z}\frac{k_x^H}{k_x^L} \tan \left( \frac{k_x^H (P_x-w_x)}{2} \right) = 0\\
\end{equation}
where $k_x^{H,L}=k(n_{H,L}^2-n_m^2)^{1/2}$, with $Z=\varepsilon_H/\varepsilon_L$ in TM and $Z=1$ in TE. Here, $\varepsilon_L=1$. Thus, when $w \ll P$, we can show that:
\begin{equation}
n_{eff,0}\approx \{
 { \begin{array}{c}
n_H [1+\frac{w}{P}(\varepsilon_H-1)]^{-1/2} \text{ in TM}, \\
n_H [1-\frac{w}{P}(\frac{\varepsilon_H-1}{\varepsilon_H}]^{1/2}  \text{ in TE},
\end{array} } 
\end{equation}
and, for $m>0$:
\begin{eqnarray}
\label{neff1D}
n_{eff,m}\approx n_H\sqrt{1-\Lambda_m^2 \left[ 1+\frac{w Z}{P-w}(\frac{\varepsilon_H-1}{\Lambda_m^2 \varepsilon_H}-1) \right]^2}\quad \\
\text{where } \Lambda_m= \frac{m\lambda}{n_H(P-w)},\quad 
\end{eqnarray}
\textit{For sake of simplicity, we can consider} $n_m\approx n_H \sqrt{1-\Lambda_m^2}$ whatever $m$. Fig.\ref{figure3}(b) gives the exact solutions taking the example of Fig.\ref{figure2}(c).

The \textit{fundamental} FP resonances will be given by the $A_m$ coefficients (when $A_m$ modulus becomes high, $B_m$ is rather negligible and reversely). The $m^{th}$ waveguide mode resonates when the real part of the determinant of $A_m$ cancels, whereas the imaginary part gives what limits the resonance (Q-factor). We see in Eq.(\ref{Am}) that, when $\lambda>P$, only $\beta_0$ is real (the zero-order diffraction beam is the only one radiating light), and a resonance may occur provided that the effective index $n_m>0$, leading then to the maximum amplitude:
\begin{equation}
A_m^{res}= \frac{i<F_m|1>}{n_{m}\sin(kh n_{m})|<1|F_m>|^2}.
\end{equation}

Given the profiles of $F_m$ when w is small (Fig.\ref{figure3}(a) and (b)), we predict a clear difference between the fundamental (gently varying) mode $m=0$, and the higher order (oscillating) ones: when $m>0$, $<1|F_m>$ is rather weak ($\gtrsim w/P$), whereas $<1|F_0>$ is of the order of unity (meaning a significant coupling with the zero-order diffracted beam). Thus, the fundamental mode does not really resonate and behaves as a background mode, with an effective index close to the material optical index (as shown in Fig.\ref{figure3}(b)). It is actually associated to classical half-wave plate conditions of bad quality factor, that slowly modulates the far-field response. \textit{We can thus be only interested in modes $m>0$ for which sharp resonances may occur}. After some calculations (see Appendix B, section \ref{appendixB}), we can obtain a practical expression:
\begin{eqnarray}
A_{m>0} (k) \approx \frac{w/P}{\sin(kh n_{m})(\Omega_m-i \Gamma_m)}\quad \quad\\
\text{where }
\label{omega} \Omega_m \approx \cot(kh n_{m}) W_m - n_m \frac{kP}{\pi}(\frac{w}{P}+\frac{P-w}{2PZ})^2\quad \quad \\
\text{and }
\label{gamma}\Gamma_m \gtrsim n_{m} (\frac{w}{P})^2 \quad \quad
\end{eqnarray}
(a more precise expression of $\Gamma_m$ can be obtained thanks to Eq.(\ref{1Fmneq0})).
These analytic results hold for a bar grating at $\theta=0^{o}$, with  $w \ll P $ and P sufficiently lower than $\lambda$ (we recall that $n_m>0$). At the first order (see section \ref{appendixB}), we find that the resonance wavelengths ($\Omega_m=0$) $\lambda_{res} \propto 4 n_m h$, and obey the relation: 
\begin{equation}\label{lres}
\lambda_{m>0,res} \approx \frac{4n_H h'} {(2p+1)\sqrt{ 1+ (m\frac{4h'}{P-w})^2}}, \end{equation}
where $h'=h+(P/\pi W_m)(w/P+(P-w)/(2PZ))^2$ is a polarization-dependent effective thickness that barely varies with $\lambda$, and $p$ a positive integer ($p=0$ corresponds to the fundamental resonance of the $m$-mode, and $p>0$ to harmonic resonances for thicker structures). Following the same development, we could find that the $B_m$ coefficients (\ref{Bm}) resonate when $\lambda \sim 4h n_m/(2p)$, so with the above result, the general law is $\lambda \sim 4h n_m/p$. ($h$ is the half-thickness of the slab). Numerical application of (\ref{lres}) in case of Fig.\ref{figure2}(c), taking $W_m\approx 0.5$: $h'=51$nm in TM and $109$nm in TE, so that we find $\lambda_{TE_1}=844$nm, $\lambda_{TE_2}=537$nm, $\lambda_{TM_1}=588$nm, and $\lambda_{TM_2}=476$nm.

$\Gamma_m$, corresponding to radiation leaks, is small (Q high) when the aperture $w$ and/or the index $n_m$ is small. However, near a cut-off mode condition, the thickness $h$ for which a resonance may occur, dramatically increases, which  necessitates very thick gratings. Regarding $\Gamma_m$, there is no special advantage to work with very high dielectric indices to get strong resonances but high optical indices comfortably widen the spectral range $[P, n_H(P-w)]$ in which fundamental sharp resonances occur. If the material becomes slightly absorbing, $n_m$ develops a small imaginary part. For example, by writing $n_m=n_a+in_b$, we find, for the first TM resonance:
\begin{equation}
\Gamma_m \sim n_a (\frac{w}{P})^2+\frac{n_b}{2n_a}[\pi W_m+  \frac{P}{h}(\frac{w}{P}+\frac{P-w}{2P|\varepsilon_H|})^2]
\end{equation}
By contrast with a moderate-Q-factor resonance, a high-quality resonance, which is bounded by few radiation leaks, will be more impacted by the introduction of dissipations, which is confirmed through Fig.\ref{figure2} when GaP becomes absorbing. A nearly cut-off mode ($n_a \rightarrow 0$, thick gratings) cannot lead to best Q-factors anymore.

At the fundamental FP resonance of a TM-mode, according to Eq.(\ref{phiy}) and (\ref{phix}), the magnetic component ($\phi_y=H_y$) shows field loops along the median plane of the slab ($z=0$) whereas $\phi_x=E_x$ is maximum near horizontal interfaces. Reversely, at the resonance of a TE-mode, $\phi_y=E_y$ meets maximum values when $z=0$ while $H_x(z=0)=0$.
For instance, taking (\ref{gamma}), the maximum electric field amplitude of the $m^{th}$ mode, inside the cavity, is bounded:
\begin{eqnarray}
|E_y(z=0)| \lesssim \frac{P}{w n_m |\sin (kh n_m)|} \text{ in TE}, \\
|E_x(z=h)| \lesssim \frac{P}{w} \text{ in TM}.\label{eq:Exmax}
\end{eqnarray}

All the expressions above are consistent with the numerical observations. Using open or bottom-closed cavities may lead to enhancements a little bit different: \textit{closed} cavities support higher amplitudes in TM because we suppress a strong electric field momentum (causing radiation) at one of the cavity outputs. Surprisingly, in TM case, the value $E_x$ at the mouth of the cavity is of the same order as that obtained with plasmonic sub-wavelength slit gratings (taking $\varepsilon_H \ll 0$). The essential difference is that it is the fundamental mode which resonates for metals, whereas it is (at least) a higher-order waveguide mode which resonates for dielectrics. Let us also recall that, from a quantum point of view, the eigen-value equation in TE-polarization is totally analogous to that of the electrons in a periodic potential $U(x)$ whose expression would be $U(x)=-[(\hbar k)^2/2M]\varepsilon(x)$, with M the mass of the electron. The energy of the wave functions $F_m$ would be then $E_m=-(\hbar kn_m)^2/2M$: high-index regions are equivalent a quantum wells for photons, whereas metals play the role of high potential barriers. 

Besides, we have observed in Fig.\ref{figure2} that the local field exhibits a (zero) dip near each FP resonance peak. Indeed, take, for instance, the field along the plane $z=0$, in the cavity:
\begin{equation}
\phi_y(x=0,z=0)= \sum_m F_m(0) A_m.
\end{equation}
Since the coefficient $A_0$ of the fundamental mode is slowly varying, whereas $A_1$ rapidly changes its phase (or sign) when going throughout the resonance spectral window, the total field is supposed to be nearly zero when $|A_1|\sim |A_0|$ on one side of the peak. This is \textit{what} gives a characteristic  Fano lineshape to the resonances we are studying here, whether in 1D or 2D. 

As a last remark, a way to minimize or even kill radiation leak of the mode $m=0$ exists, by exciting the  grating with an \textit{evanescent} wave, so that $\beta_0$ becomes also imaginary in Eq.(\ref{Am}) and (\ref{Bm}). This effect has been shown for metallic gratings \cite{Quemerais,leperchec2010}.

\subsection{2D grating (TM case)}

We still consider normal incidence and, to be more concise, discuss only the case of the TM-excitation defined in Fig.\ref{figure1}. From Maxwell equations and variables separation, we find that each field component is a linear combination of eigen functions $F^{(m,l)}$ whose behaviours, in the high-index regions (excepting at cavity corners), are:
\begin{eqnarray} \label{eqHy}
H_y^{(m,l)} (x,y,z) \sim \cos (k_x^H x)\cos (k_y^H y )e^{\pm i k_z z}\quad \quad \\
\label{eqEx}
E_x^{(m,l)} (x,y,z) \sim \frac{k_z^2+(k_y^H)^2}{k\varepsilon_H k_z}\cos (k_x^H x)\cos (k_y^H y )e^{\pm ik_z z},\quad \quad\\
E_y^{(m,l)} (x,y,z)\sim \frac{k_x^H k_y^H}{k\varepsilon_H k_z}\sin (k_x^H x)\sin (k_y^H y )e^{\pm i k_z z},\quad \quad\\
E_z^{(m,l)} (x,y,z)\sim \frac{k_x^H}{k\varepsilon_H}\sin (k_x^H x)\cos (k_y^H y )e^{\pm i k_z z},\quad \quad \label{eqEz}
\end{eqnarray}
$k_x,k_y,k_z$ implicitly depend on mode indices $(m,l)$ (see hereafter), with $k_z^2+(k^H_x)^2+(k^H_y)^2 =k^2\varepsilon_H$, and $k_z^H=k_z^L$ whatever $x,y$ within the grating. (By putting $k_y=0$, we retrieve the 1D case). In TM, $H_x$ is nearly zero, whereas $H_y$ and $E_x$ are the even functions. Also, $H_y$ is continuous everywhere and almost constant inside the cavity. (An other similar set of equations can be obtained for TE-modes, for which by $H_y=0$). 

We are interested in vertical resonances of modes whose effective indices $n_{eff}=k_z/k \in [0,n_H]$ ($\lambda_{FP}\propto 4n_{eff}h$). The resolution of the exact eigenvalue problem of 2D gratings is a difficult issue and the scientific literature gives very few works about this (excepting the Fourier modal expansion method \cite{Li2014,Attiya}, called RCWA). There is no simple transcendental equation as in 1D. We can find in \cite{Sharma2010,Yeap} some detailed procedures for a rectangular waveguide, to which bi-periodicity conditions should be added. Instead, to find an approximate  expression of $n_{eff}$, we assume the 2D void \textit{locally} results from a combination of two crossed (periodic) 1D trenches. 
We are driven to solve two transcendental equations (see (\ref{eqtranscendentalTMl})), one equivalent to TM-polarization for $k_x$, and one equivalent to TE-polarization for $k_y$.  
As $w_{x,y} \ll P_{x,y} <\lambda$, the hole is only perturbative so that $k^H_{x} \approx 2m\pi/(P_x-w_x)$ and $k^H_{y} \approx 2l\pi/(P_y-w_y)$, where the integers $m,l \ge 0$. Therefore,
\begin{equation}\label{neff2D}
n_{eff}^{(m,l)} \sim n_H \sqrt{1-\left(\frac{m\lambda}{n_H(P_x-w_x)}  \right)^2 - \left(\frac{l\lambda}{n_H(P_y-w_y)}  \right)^2 }.
\end{equation}
A more precise expression could be written by using (\ref{neff1D}) (in TM for $k_x$ calculation, and TE for $k_y$). This is to be compared to the simple \textit{perfect metal} case where the indices of the cavity modes are $n_{eff}^{(m,l)}=[1-(\frac{m\lambda}{2w_x})^2-(\frac{l\lambda}{2w_y})^2]^{1/2}$, as is well known in the theory of waveguides and resonators. However, in a hollow rectangular waveguide, the fundamental mode giving a non-null field solution is $(0,1)$ (taking $w_x\leq w_y$), so for very small square voids, all the modes are strongly evanescent and cannot lead to FP-type resonances. Let us come back to dielectrics. We still put aside the $(0,0)$-mode as strong resonances only occur for higher orders. According to Eq.(\ref{neff2D}), one can guess the spectral range where a mode is propagating or cut-off depending on the ratios $\lambda/(n_H(P-w))$. For instance, no $(1,1)$-mode resonance is expected when $\lambda>800$nm if $P_x=P_y=350$nm, whatever $h$. In the reference example of Fig.\ref{figure2}(a) and (b), around $\lambda_{TM(1,1)}=556$nm,  Eq.(\ref{neff2D}) gives $n_{eff}=2.54$ with $w=11$nm, and $n_{eff}=2.36$ with $w=35$nm, whereas $n_{eff,RCWA}\approx 2.6$ for both cases. 

On the basis of the analytic calculation performed in 1D (see Eq.(\ref{lres})), the resonance wavelengths can be estimated as:
\begin{equation}\label{lres2D}
\lambda_{res}^{(m,l)}\approx \frac{4n_H h'} {p\sqrt{ 1+ (m\frac{4h'}{P_x-w_x})^2 + (l\frac{4h'}{P_y-w_y})^2}},
\end{equation}
with $p \geq 0$, and the \textit{maximum electric field amplitude} (at the hottest spot, inside the nanovoids) \textit{will be directly bounded by the surface ratio} $[w_xw_y/(P_xP_y)]^{-1}$. As in Eq.(\ref{eq:Exmax}), this quantity naturally comes from the overlap integral between the zero-order Rayleigh wave $e_{0,0}=e^{ik(\gamma_0 x+\gamma_0 y)}=1$ and the even eigen mode, which should appear in the denominator of the coefficients. 
A numerical application of (\ref{lres2D}), to be compared to Fig.\ref{figure2}(b), gives: $\lambda_{TM_{11}}=555$nm with $R=(P/w)^2=1000$, and $520$nm with $R=100$. 

\begin{figure}
\begin{center}
\includegraphics[angle=0,scale=.5]{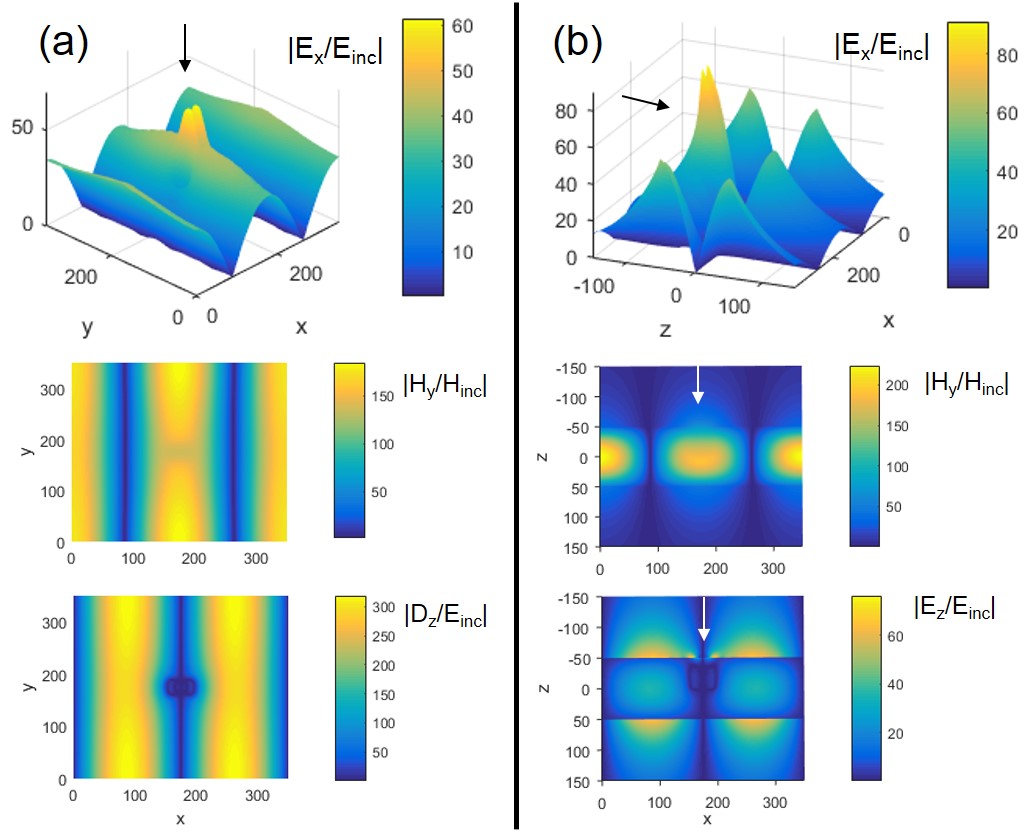}
\end{center}
\caption{$TM_{1,0}$ resonant mode at $\lambda=624.33$nm when $w=35$nm (see Fig.~\ref{figure2}(b)). $n^{(1,0)}_{eff} = 2.8$ (RCWA value). (a) Maps of the main electromagnetic components in the $(x,y)$-plane, at $z=-h/2$ (void at the center of the maps). $D_z=\varepsilon(x,y) E_z$ is the electric displacement vector. Arrows indicate the direction of the incident wave. $H_x$ and $E_y$ are negligible. (b)  Maps following the $(x,z)$-cross-section at $y=P_y/2$. Contrary to $D_z$, $E_z$ is not continuous at horizontal interfaces and better highlights the field enhancement above the free surface. }\label{figure4}
\end{figure}


Same calculation could be done for TE-modes, and Eq. (\ref{lres2D}) gives $\lambda_{TE_{11}}=684$nm with $R=1000$, and $660$nm with $R=100$. Even if some indices $n_{eff}$ are close (with possibly degenerate modes, especially with a square geometry), TM or TE resonance wavelengths are not necessarily the same because of different phase conditions. 
As said for 1D gratings (see Eq.(\ref{lres})), a $TE_{0,1}$ resonance always occurs at a frequency \textit{lower} than that of a $TM_{0,1}$ resonance. A complete 2D modal analysis would actually deserve a full paper, we will not give further details here in order to focus us on numerical results in the following.

\section{Electromagnetic field maps}
\label{section:maps}

\begin{figure}
\begin{center}
\includegraphics[angle=0,scale=.47]{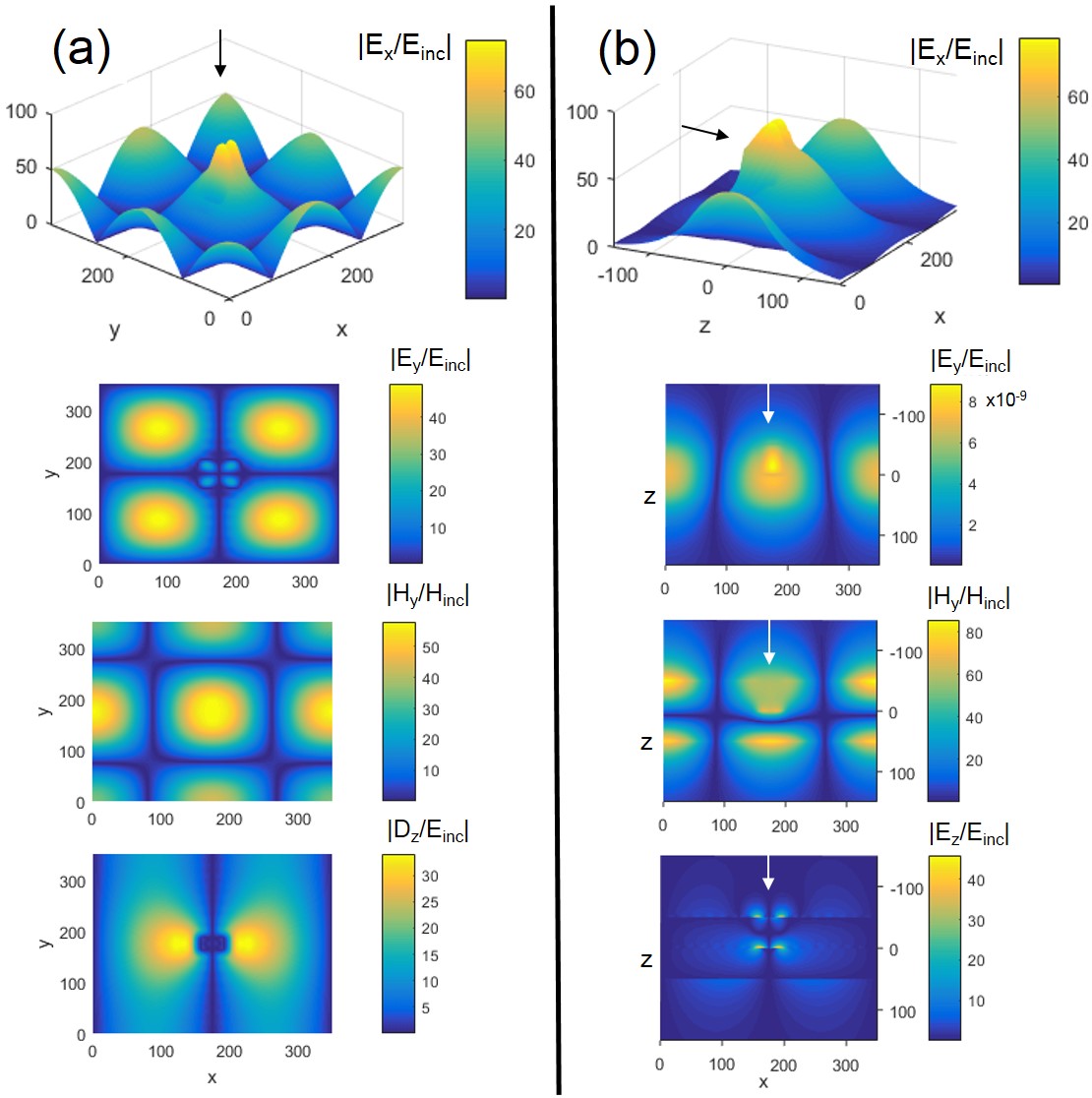} \end{center}
\caption{FP resonance of the quasi-$TE_{1,1}$ mode at $\lambda=656.76 nm$ for the case $w=35nm$ (see Fig.~\ref{figure2}(b)). (a) Maps of four electromagnetic components in the $(x,y)$ plane at $z=-h/2$ (mid-height of the voids), with $D_z=\varepsilon (x,y) E_z$. (b) Field maps according the $(x,z)$ cross-section at $y=P_y/2$. Arrows indicate the incidence direction. This is actually a hybrid TM/TE mode: in case of a pure $TE_{1,1}$ mode (true TE-polarized excitation), $E_x$ and $E_y$ would exchange their $(x,y)$-distributions, $E_y$ would keep the \textit{same} type of profile in the $(x,z)$-cross-section but would support, this time, a strong amplitude, especially inside the void.}\label{figure5}
\end{figure}

At the FP resonance of a TM-mode, the magnetic field $H_y$ exhibits maximum amplitude along the median plane of the slab ($z=0$) whereas $E_x$ is maximum near horizontal interfaces. Reversely, at the FP resonance of a TE-type mode, the plane $z=0$ is the one where $E_x$ or $E_y$ have maximum values while it is nodal for $H_y$ and $H_x$.
These behaviours are the same for 1D or 2D. This is shown in Fig.\ref{figure4} and \ref{figure5} for the intermediate aperture case of Fig.\ref{figure2}(a) and (b), i.e. $w=35$nm ($R=100$).

Figure \ref{figure4} gathers the most relevant field maps at the $TM_{1,0}$ resonance. The cavity is clearly the site of a significantly enhanced electric field modulus, compared to the high-index region, with a maximum value near the mouth of the nano-pit. The whole horizontal surfaces also support a huge normal field $E_z$  (intensity $\sim 5.10^3$). The latter waves are evanescent on both parts of the slab and do not radiate energy. Consistently with the preceding section, the radiation leakage mainly comes from the hole aperture, where the tangential $E_x$ momentum behaves as a localized oscillating dipole, very reminiscent of what we have with 1D plasmonic cavities \cite{leperchec06,leperchec15}. 

Figure \ref{figure5} shows different field maps at a quasi-$TE_{1,1}$ resonance. The nano-void is still the hot spot of the structure but, this time, the maximum electric field enhancement is obtained at the \textit{bottom} of the pit ($z=0$). Also,  the electromagnetic patterns on the horizontal free surface are different (impossible in 1D), with four-lobe distributions in the $(x,y)$-plane and an overall weak $E_z$ component.  Note that the field patterns and the relative amplitude of each component, in the numerically computed maps of Fig.\ref{figure4} and \ref{figure5}, very well reflect the modal expressions given by (\ref{eqHy}) to (\ref{eqEz}). These last equations are sufficiently powerful to predict some field configurations, except at the corners of the cavities where the fields may show localized peaks of moderate amplitude. 

Even higher amplifications would be observed in the case $w=11$nm and could not occur within equivalent square metallic voids because the localized cavity mode (fundamental mode) would be cut-off when $w_x,w_y$ are both lower than $\lambda/3$ to $\lambda/4$ typically, depending on the hole aspect-ratio and the metal permittivity \cite{Gordon}. If light-trapping in metallic nanovoids is possible, based on collective or individual resonances, it rather occurs at shorter (near-UV) wavelengths \cite{Dunbar}. 

Such sensitive transparent substrates are obviously relevant for molecule detection or fluorescence enhancement \cite{Ganesh}, and strong localized optical forces should occur, like for metallic cavity gratings \cite{Velzen}. With an appropriate sizing, the strong electric field may be used to boost non-linear effects, especially for second harmonic generation using the large $\chi^{(2)}$ non-linearity of GaP here, as demonstrated with a photonic crystal cavity \cite{Rivoire}.

It is worth pointing out that a periodic layer with cubic voids \textit{entirely} immersed within the medium may also resonate efficiently (not shown here). In this case, the excitation and the weak radiation leak should occur through direct coupling between the resonant mode and (at least) two first Rayleigh orders which are propagating inside the homogeneous sub-layers of the slab (one of both orders being internally trapped).

A general comment: to get potentially very high Q-factors, a challenge is to technologically make structures with minimum imperfections (surface roughness) which would yield scattering losses \cite{Zhou}, especially at horizontal interfaces. It also implies that we work with sufficiently extended gratings, with no high-index substrate too close to it, which could, otherwise, convert the surrounding evanescent field, into radiation leaks \cite{Sadrieva}. However, a substrate underneath may be used to get sensors with positive transmission signals \cite{Chang2018}. Besides, to get more compact or miniaturized  devices, the integration of the grating in a cavity-resonator may be an interesting solution \cite{Kintaka}. 
By illuminating the structure with a light cone, we may willingly widen the resonance spectral width. 

\section{Multi-cavity and / or thicker gratings: frequency-susceptibility boosting}
\label{section:freqsusceptibility}
\begin{figure}
\begin{center}
\includegraphics[angle=0,scale=.46]{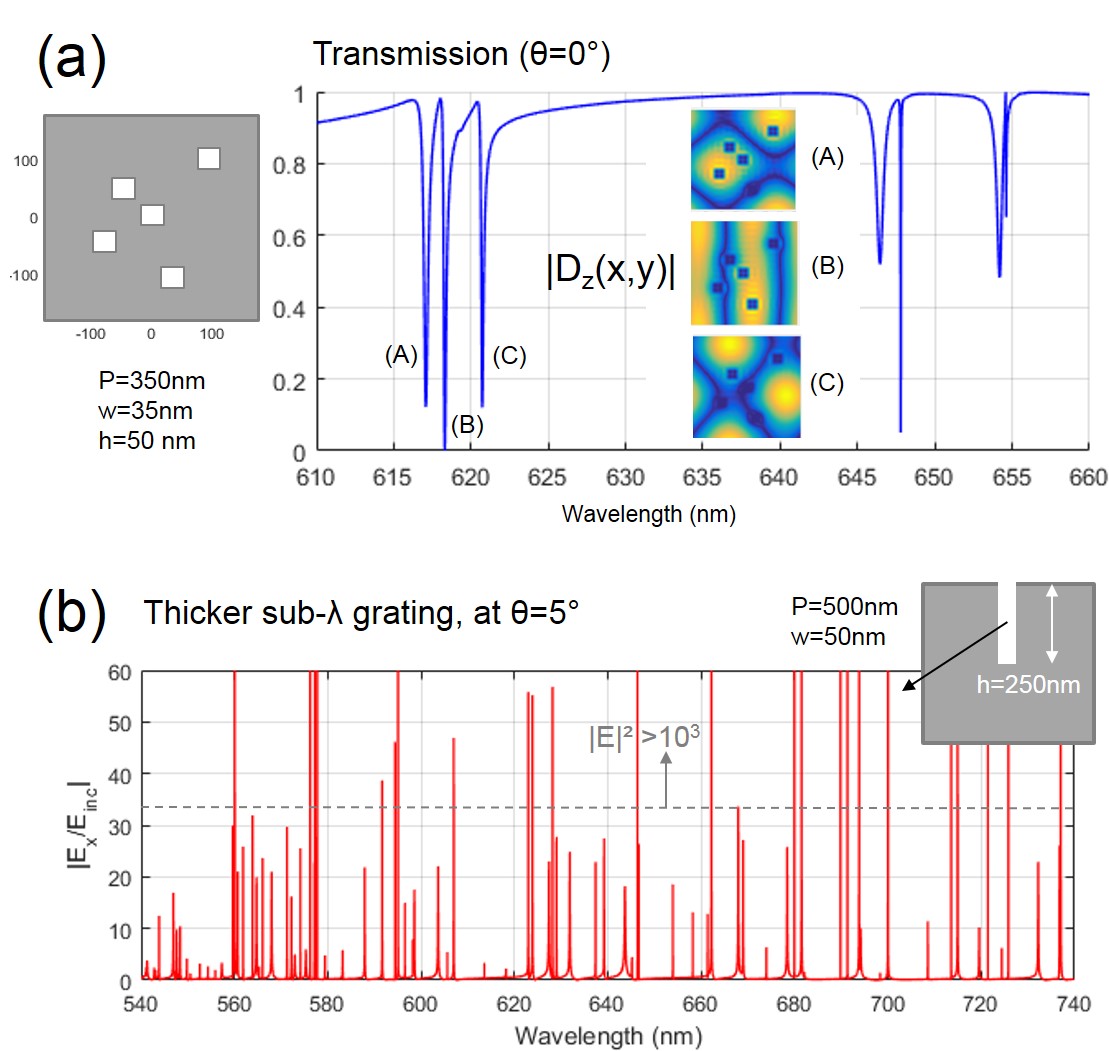}
\end{center}
\caption{(a) Optical response of a multi-void irregular grating (sketch given on the left), at $\theta=0^{o}$, in the spectral window where one TM resonance and one TE resonance occur for the single-void case (see Fig.\ref{figure2}(a)). Small inserted maps of the electric field induction correspond to the three resonance dips (A), (B), (C). (b) Broadband frequency susceptibility of a widened square 2D grating with a single pit per period, under oblique incidence ($\theta=5^{o}$). Electric field enhancement calculated at mid-height of the voids. Scale is limited to 60 in order to exhibit numerous minor peaks (some peaks overpass 100).}\label{figure6}
\end{figure}
Sharp resonances do not necessarily rely on an ideal or simple geometry. They may indeed persist when other neighboring dielectric cavities are inserted inside the unitary periodic cell. Figure \ref{figure6}(a) gives the response of a grating containing five voids of same size, but randomly-distributed, to be compared to the simple grating already shown in Fig.\ref{figure2}(a). The multi-void case gives new resonance peaks. \textit{Odd} eigen modes or new hybrid TE/TM ones may indeed be excited at normal incidence because of the structural symmetry breaking. For instance, the TM resonance around $620$nm, splits in different variants, with the same type of field profile in the $(x,z)$-plane, but sometimes crossed waveguide modes in the $(x,y)$-plane, as illustrated by the electric induction maps. The average amplitudes are a bit lower at the surface level ($|E_z|\sim 40$) because of a greater effective aperture. 
These new sets of electromagnetic field distribution, due to mixing of waveguide modes, is a phenomenon different from what happens with multi-groove metallic gratings, where it is the dipolar coupling between adjacent resonators which leads to contrasted light localizations \cite{Skigin,Barbara2009}. 

Let us come back to the one-cavity grating. A way to multiply the number of resonances, in a given spectral window, is the use of a thicker dielectric slab and/or a larger bi-periodicity, which introduces harmonic or other higher-order modes. Another possibility is to resort to oblique incidence because odd eigen modes will also be excited. By combining the two latter ways, we can get a structure showing an impressive frequency susceptibility, both for far-field and near-field, although \textit{all} its geometrical parameters remain sub-wavelength. Figure \ref{figure6}(b) illustrates this point for a square grating $500$nm thick, with a period $P=500$nm (still sub-$\lambda$), and $R=100$. A minimum wavelength resolution of $0.01$nm is needed to not to overlook some discrete details and abrupt variations (a smaller $w$ would have led to even sharper variations). Numerous peaks are thus predicted in a range going from green to near-infrared wavelengths, with up to fourteen powerful resonances in the tight $[560,580]$nm range (around one resonance every $1.5$nm wavelength). The normalized intensities $|E|^2$ inside the voids are quite often greater than three orders of magnitude. Other light amplifications may also occur at different locations of the structure. Transmission (not shown) is impacted by a strong concentration of coupled dips and peaks, i.e. we can meet zero or perfect transmission at many specific wavelengths (and conversely for the reflectivity). 

By mixing larger geometrical parameters and/or other incidence conditions and/or multi-cavity arrangements, we might almost create a quasi-continuum of high-Q resonances, increasing then the probability of occurrence of field effects needed for applications like the ones already cited before. Very thick corrugated gratings (with holes or narrow grooves) could be made by resorting to deep trench isolation techniques, well known in silicon microelectronics \cite{Tournier}. 

\section{Ultra-thin membranes: filling space with high amplified fields}
\label{section:ultrathin}

\begin{figure}
\begin{center}
\includegraphics[angle=0,scale=.55]{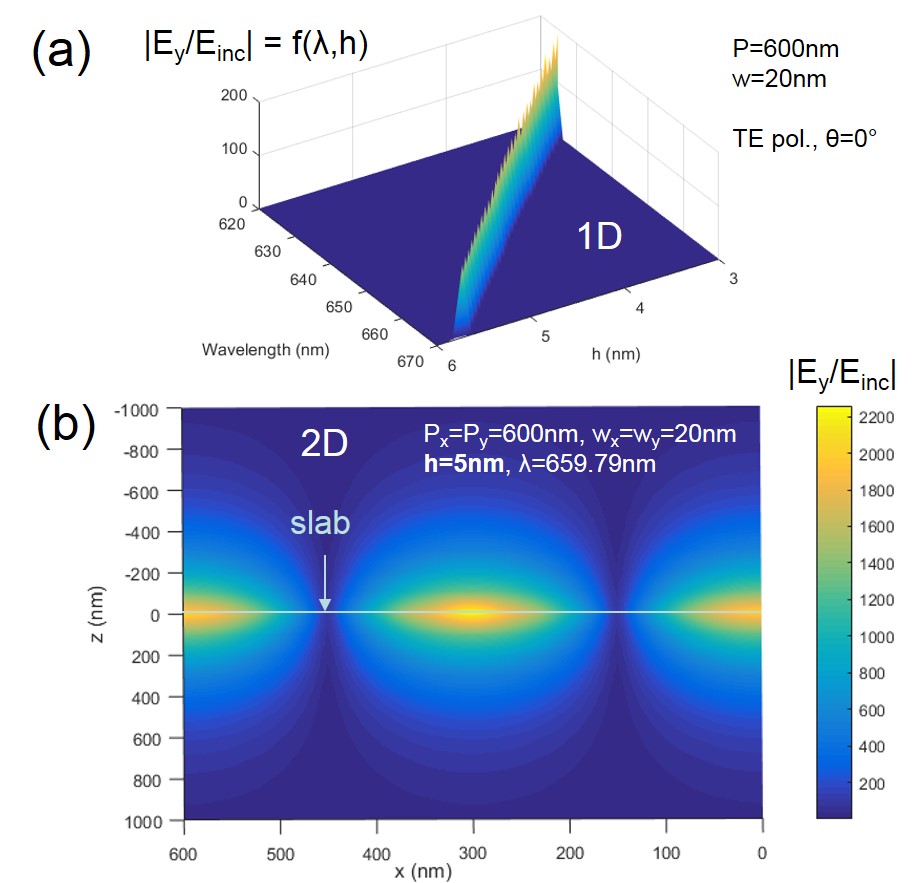}
\end{center}
\caption{(a) Diagram of the fundamental resonance of the $TE_1$ mode in function of $(\lambda,h)$ of a 1D grating taking $P=600$nm and $w=20$nm. We assume that permittivity is valid for ultra-small thickness. (b) Electric field maps of the equivalent 2D dielectric grating at the fundamental resonance, under TE polarization, when $h=5$nm only (normal incidence).}\label{figure7}
\end{figure}

We now explore the case of a very thin slab and focus us on the most sub-wavelength resonance, namely the fundamental $ TE_{0,1}$ resonance, with maximum enhancements along the plane $z=0$. Take again the general equation (\ref{Am}), from the 1D study. We consider here that $\lambda$ and $P$ may be neighboring. Let us assume that the higher order coupling terms $|q|\geq 2$, in the sum of the denominator, are not preponderant. Then, if $h \ll \lambda$, we immediately get:
 \begin{eqnarray}
A_m \approx \frac{<F_m|e_0>}{\Omega^* -i\Gamma^*}, \quad \\
\label{omegastar} \text{with : } \Omega_m^*=W_m - \frac{2 k h n^2_{m}}{\sqrt{(\lambda/P)^2-1}}  |<e_1|F_m>|^2, \quad \\
\label{gammastar} \text{and  }\Gamma_m^*= k h n^2_{m}  |<e_0|F_m>|^2.\quad 
\end{eqnarray}

Therefore, one analytically sees that the resonance $\Omega^*=0$ can indeed \textit{persist for very small $h$ provided $\lambda$ is close to P}, and the resonance wavelength will then follow a law $\lambda_{res} = P\sqrt{1+O((kh)^2)}$. This is exactly what we get through full-wave simulations, assuming that the dielectric permittivity is still valid for thin material volumes: Figure \ref{figure7}(a) displays the resonance diagram of a 1D structure under TE-polarization, in function of $(\lambda,h)$ for a given period $P=600$nm and $w=20$nm (P is chosen so that the resonance remains in the range where GaP is not absorbing). 

We can retrieve the same behaviour for the equivalent bi-periodic structure. By taking $h=5nm$ only (see Fig.\ref{figure7}(b)), the light intensity enhancement may reach a remarkable value $|E_y/E_{inc}|^2\approx 5.10^6$ at $\lambda=659.79$nm (the resonance strength is much greater than in 1D). There is no more vertical FP character, but a Fano resonance rather governed by the lateral dimensions (Bragg regime). Of course, if $h=0$, $\Omega^*$ cannot cancel and we loose any resonance possibility. Also, if $n_m$ becomes imaginary (mode cut-off), $\Omega^*$ cannot cancel anymore, so it is mandatory to maintain $\lambda<n_H(P-w)$, according to (\ref{neff1D}). The limiting factor $\Gamma^*$ is linearly proportional to the thickness, and for a given $h$, the resonance amplitude inversely increases with a smaller $w$.

By taking an extreme set of parameters ($ h\rightarrow 0$, $\lambda \rightarrow P^+$), according to Eq.(\ref{omegastar}) and (\ref{gammastar}), we might theoretically reach an \textit{opto-geometrical catastrophe}, with tremendous light amplifications on a dielectric membrane containing nano-pinholes. The Q-factors could become so high that it would require perfect structures (without surface irregularity or material impurity), or would be experimentally imperceptible. For reason of mechanical robustness, surrounding the membrane by a low index material (like $SiO_2$) is well possible, with a period lower than in the vacuum-case in order to keep a zero-order diffraction grating.

Finally, the reactive power, which is a quantity of energy stored by the system (in permanent regime) and proportional to the Q-factor, flows at the external surface as evanescent waves, according to the following expression, in 1D \cite{leperchec2010}:
\begin{equation} 
ReactP(z>h)=\Im [\sum_q \beta_q |T_q|^2 e^{-2(z-h)\Im(\beta_q)} ],
\end{equation}
where $ReactP$ is normalized to the incident power flux, over a whole period. The terms $T_q$ are a linear combination of the coefficients $A_m$ and $B_m$.  When $\lambda \rightarrow P^+$, $\beta_1 \rightarrow i0^-$. Consequently, for these huge resonances, we have also a maximum extension of the field in the space surrounding the structure. Through Fig.\ref{figure7}(b), we indeed confirm that the field moderately decreases on both sides of the grating and, even at $z=1\mu$m ($=1.5\times\lambda$), the normalized intensity still keeps a very high value ($10^3$). 

It would be a challenge to carry out an experiment demonstrating these predictions on ultra-thin holed slabs, in spite of likely difficulties given imperfections of real structures. Nevertheless, measurements of Q-factor as high as $10^6$, like that of bound states in the continuum, have already been reported \cite{Hsu2013}. 

\section{Conclusion}

This theoretical work aims at giving, with the support of numerical and analytic means, a simple comprehensive vision of a high-dielectric-index plate containing very small rectangular pits, with a special attention given to the Fabry-Perot resonances of high-order eigen modes and their local \textit{electric}-field enhancements. As a reference example, we took the GaP waveguide slab under visible excitation.
We have resorted to a true modal method as a powerful tool to generate explicit parametric expressions allowing to interpret the behaviors numerically observed in 2D (at normal incidence). For that, we have used a weak indirect-coupling condition between grating eigen-modes. This analytic development should help opticians to rapidly predict the resonance rules of these canonical photonic structures and, therefore, to design devices for applications from ultraviolet to THz.  
At frequencies lower than visible ones, it should be easier to make strongly sub-wavelength voids and, then, to get extreme phenomena.

To summarize, we have underlined the central link between the strength of the high-Q-factor resonances and the ratio between the aperture area of the pits and the periodic surface $P_xP_y$. The cavity is both the light scatterer towards the vertically propagating grating modes but also the site where a field component may locally couple with the external diffraction order (hence radiation leaks). Elsewhere, trapped waves give an evanescent field on the horizontal interfaces. For that reason, a 2D structure allows field enhancements markedly greater than in 1D. Depending on the TM/TE excited modes, field maps show a significantly boosted electrical component inside the nanopits, either at the bottom, either at the mouth, and sometimes a strong electric field over the whole surface. The identification of these different configurations should guide the choice of the right mode selection for light trapping, molecule spectroscopy or to maximize some non-linear effects, for instance. (Nevertheless, further theoretical investigations could well be carried out, for instance at conical incidence, with new dispersion diagrams).
As explained, hot spots based on FP mechanisms cannot be obtained in square nano-holes made in a metallic screen for same excitation frequencies. Also, an irregular (multi-cavity) dielectric grating does not preclude sharp resonances, and may give rise to modal variants. We pointed out that resizing appropriately the grating, and switching on additional odd eigen modes through oblique incidence, we can increase the spectral concentration of resonance peaks, leading to a strong frequency-susceptibility of the sub-wavelength patterned surface.
We finally saw that light can be caught by an ultra-thin membrane (a few nanometers thick) while the wavelength is superior to the periodicity. This Fano resonance is associated to a maximum extension of a non-radiative field on both parts of the slab, making the surrounding space as a (reactive) energy tank with giant electric field intensities.


\small
\section{Appendix A: Weak indirect-coupling condition in the exact modal method (1D)}\label{appendixA}

The true eigen modes form  a rigorously orthogonal basis  so that there no direct coupling between them, and the intrinsic resonance characteristics of the $m^{th}$ mode (i.e. the determinants of $A_m$ and $B_m$) do \textit{not} depend on the other modes. However, the excitation strength (i.e. the numerators of $A_m$ and $B_m$) is possibly influenced by the other modes, due to indirect-coupling terms \textit{via} the external diffracted field (see Eq.(\ref{Xmp})). So, to get  the coefficients $A_m$ and $B_m$  in (\ref{Am}) and (\ref{Bm}), from the exact matrix results \cite{Botten1981}, we use a weak indirect-coupling condition for all eigen modes, which is like using a kind of mono-modal expansion  (inside the grating only) for each mode.
Eq.(\ref{Am}) and (\ref{Bm}) are then quasi exact when the mode is supposed to support most of the electromagnetic power. When a mode \textit{badly} resonates or when several eigen modes resonate in a tight spectral range,  expressions (\ref{Am}) and (\ref{Bm}) are less valid and rigorous, regarding the numerators, since the energy is further redistributed between each excited modes.

To better understand what happens in the calculation, let us restrict the matrix solution of the 1D exact modal method to the two first eigen modes. We consider only dielectrics at normal incidence and thus only even eigen functions (it can be shown that the overlap integrals $<F_m|e_q>$ defined in Eq.(\ref{coupling}) must all have real values). We can therefore write:
\[ M \cdot \left(\begin{array}{c} A_0 \\ A_1 \\ B_0 \\ B_1 \end{array}\right) =\left(\begin{array}{c} 2 <F_0|e_0>\\ 2 <F_0|e_0> \\ 0 \\ 0 \end{array}\right)  \]
with M a $4\times 4$ matrix as follows:
\begin{equation} \label{eqmat}
   M=
  \left[ {\begin{array}{cccc}
   C_0W_0+X_{0,0}S_0 ; X_{0,1}S_1 ; -S_0W_0+X_{0,0}C_0 ; X_{0,1}C_1 \\
    X_{1,0}S_0 ; C_1W_1+X_{1,1}S_1 ; X_{1,0}C_0 ; -S_1W_1+X_{1,1}C_1 \\
   C_0W_0+X_{0,0}S_0 ; X_{0,1}S_1 ; S_0W_0-X_{0,0}C_0 ; -X_{0,1}C_1 \\
    X_{1,0}S_0 ; C_1W_1+X_{1,1}S_1 ; -X_{1,0}C_0 ; S_1W_1-X_{1,1}C_1 \\
  \end{array} } \right]
\end{equation} 
where
\begin{eqnarray}  
W_m=<F_m|F_m>,\\
S_m = n_m\sin(khn_m) ,\\
C_m =n_m\cos(khn_m),\\
\label{Xmp} X_{m,p}=\sum_q \frac{<F_m|e_q><e_q|F_p>}{i\beta_q}.
\end{eqnarray} 
For instance, from (\ref{eqmat}), we find:
\begin{equation} 
B_0=\frac{(C_0 W_0+X_{0,0}S_0)A_0+ X_{0,1}(S_1 A_1 - C_1 B_1)}{-S_0 W_0+X_{0,0}C_0},
\end{equation} 
\begin{equation} 
A_0=\frac{2 <F_0|e_0>-X_{0,1}(S_1 A_1 + C_1  B_1)+(S_0 W_0-X_{0,0}C_0)B_0}{C_0 W_0+X_{0,0}S_0}.
\end{equation} 

Consequently, if we neglect the indirect-coupling terms ($X_{0,1}$), we immediately find the results (\ref{Am}) and (\ref{Bm}) for $m=0$, and $m=1$, separately. It does not change the denominators. We can extend the reasoning to the full eigen basis (same matrix structure). Generally, one of the most significant indirect-coupling terms of a given $m$-mode is the one with the fundamental mode $m=0$ (ie ($X_{0,m}$)), because of the almost non-oscillating nature of the latter (it is a kind of background mode which easily couples with the diffraction orders radiating outside). 

Let us stress that the right calculation of the far-field coefficients cannot omit the indirect-coupling terms. We recall that, according to the exact modal development:
\begin{equation} 
T_q=\frac{1}{i\beta_q}\sum_q n_m [-A_m\sin(khn_m) + B_m\cos(khn_m)]<e_q|F_m>
\end{equation} 
and the zero-order transmission is $|T_0|^2$.

\section{Appendix B: Getting simplified field coefficients and FP-resonance wavelengths}\label{appendixB}

We consider the general expression of $A_m$ in Eq.(\ref{Am}). At normal incidence, $\gamma_q=q\lambda/P$ and $\beta_q=\beta_{-q}=\sqrt{1-(q\lambda/P)^2}$. Thus,
\begin{eqnarray}
A_m= \frac{<F_m|1>}{\sin(kh n_{m})(\Omega-i \Gamma)}, \\
\Omega_m = \cot(kh n_{m}) W_m - i 2 n_m \sum_{q>0} \frac{|<e_q|F_m>|^2}{\beta_q}, \\
\Gamma_m =  n_{m}  |<1|F_m>|^2 .
\end{eqnarray}

We do not consider the fundamental grating mode $m=0$ because it does not resonate. For modes $m>0$, the eigen functions are oscillating (with nodes and anti-nodes) inside the high-index region, whereas they are nearly constant (close to unity) around the void region, according to Fig.\ref{figure3}.
Consequently, the coupling terms $<1|F_m>\gtrsim w/P$, in TM or TE-polarization.
A more precise calculation, at the first order, based on $n_m$ in (\ref{neff1D}), leads to:  
\begin{equation}\label{1Fmneq0}
<1|F_{m\neq 0}>=\frac{w}{P}+\frac{P-w}{P}\sec \left(m\pi \left|1+\frac{w Z}{P-w}(\frac{\varepsilon_H-1}{\Lambda_m^2 \varepsilon_H}-1) \right| \right)
\end{equation}
with $\varepsilon_H\Lambda_m^2 =(m\lambda/(P-w))^2$ and $Z=\varepsilon_H$ in TM, $Z=1$ in TE.
It is interesting to note that when $m\lambda=\sqrt{\varepsilon_H-1}(P-w)$, $<1|F_m>$ (i.e. radiation leaks) should be \textit{minimized}. This is also a condition at which $n_{m,TM}=n_{m,TE}$.

Things are different concerning $<e_{q\neq 0}|F_m>$ because the eigen functions and the oscillating Rayleigh waves $e_q(x)$ may sometimes strongly resemble: especially, $<e_m|F_m>$ is expected to be significant. The modal representation shows that when $w\ll P$, $F_m(x)\sim \cos(k_x x)$ for the first modes (the other modes are strongly evanescent and negligible), with $k_x\approx 2\pi m/(P-w)$ in the high-index regions. Therefore,
\begin{equation}
<e_m|F_m>=\frac{1}{P} \int_{0}^{P} \cos(2m\pi \frac{x}{P})\frac{F_m(x)}{\omega(x)}dx\approx \frac{w}{P}+\frac{P-w}{2PZ},
\end{equation}

While $P$ is sufficiently lower than $\lambda$,  $\beta_q\approx i q \lambda/P$ when $q>0$, and we get:
\begin{equation}
\sum_{q>0} \frac{|<e_q|F_m>|^2}{\beta_q}\approx \frac{P}{i\lambda}(\frac{w}{P}+\frac{P-w}{2PZ})^2 +\sum_{q>1} \frac{|<e_q|F_m>|^2}{i q\lambda/P}.
\end{equation}

By omitting the residual summation term (assuming weak overlap with higher Rayleigh orders), one finds:
\begin{equation}
\Omega_m \approx \cot(kh n_{m}) W_m - n_m \frac{kP}{\pi}(\frac{w}{P}+\frac{P-w}{2PZ})^2.
\end{equation}
The resonance will occur when $\Omega_m=0$. Let us write $\Omega_m =\cot(kh n_{m}) W_m-Y$. As $Y$ is small, we may write $kh n_{m}\approx (2p+1)\frac{\pi}{2}-Y/W_m$ at the first order, and the resonance wavelength becomes:
\begin{equation}
\lambda_{m,res} \approx \frac{4n_m(\lambda)}{(2p+1)} [h + \frac{P}{\pi W_m}(\frac{w}{P}+\frac{P-w}{2PZ})^2].
\end{equation}
We can rewrite it under the form $\lambda_{m,res} \approx 4n_m h'/(2p+1)$. Finally, by using $n_m(\lambda)\approx n_H\sqrt{1-(\frac{m\lambda}{n_H(P-w)})^2}$, one gets:
\begin{equation}
\lambda_{m,res} \approx \frac{4n_H h'} {(2p+1)\sqrt{ 1+ (m\frac{4h'}{P-w})^2}}
\end{equation}
We recall that $h$ is the half-thickness of the grating.\\

\end{document}